\documentclass{article}

\PassOptionsToPackage{numbers, compress}{natbib}

\usepackage[preprint]{neurips_2026}

\usepackage[utf8]{inputenc}    
\usepackage[T1]{fontenc}       
\usepackage{hyperref}          
\usepackage{url}               
\usepackage{booktabs}          
\usepackage{amsfonts}          
\usepackage{amsmath,amssymb}
\usepackage{nicefrac}          
\usepackage{microtype}         
\usepackage{xcolor}            
\usepackage{graphicx}
\usepackage{multirow}
\usepackage{array}
\usepackage{makecell}
\usepackage{caption}
\usepackage{subcaption}
\usepackage{enumitem}
\usepackage{xspace}

\newcommand{\sysname}{Confucius4-TTS\xspace}

\title{\sysname: Transcript-Free Cross-Lingual Zero-Shot TTS with a Learnable Speaker Encoder}

\author{%
  Huaxuan Wang \quad Huimin Wang \quad Ruiyu Zhang \quad Yingjie Li \quad Yitao Duan \\
  NetEase Youdao, Beijing, China\\
  \texttt{\{wanghx04, wanghm08, zhangry05, liyj, duan\}@rd.netease.com}
}

\begin{document}

\maketitle

\begin{abstract}
Recent advances in zero-shot text-to-speech (TTS) have substantially improved speech quality and voice cloning fidelity. However, many zero-shot TTS systems still depend on audio prompt transcripts at inference time. This dependency limits cross-lingual voice cloning, since in-the-wild reference audio is often untranscribed. In this technical report, we present \sysname{}, a multilingual zero-shot TTS system that supports 14 languages and performs both intra-lingual and cross-lingual reference cloning without requiring transcripts of audio prompts. \sysname{} follows a two-stage architecture, consisting of text-to-semantic (T2S) and semantic-to-acoustic (S2A) modules. The LLM-based T2S module uses a learnable speaker encoder to extract timbre features from self-supervised speech representations, and the conditional flow-matching S2A module converts the predicted semantic tokens into mel-spectrograms. The same model also supports continuation cloning when a reference transcript is available. \sysname{} is trained on large-scale multilingual speech data. It achieves high intelligibility and speaker similarity on public benchmarks. On the CV3-Eval cross-lingual benchmark, \sysname{} obtains an average WER of 3.73\% across six directions. On our internal cross-lingual set, it achieves the best average overall rank in human evaluation among recent open-source and commercial systems. We release code, model checkpoints, and demos at \url{https://github.com/netease-youdao/Confucius4-TTS}.

\end{abstract}

\section{Introduction}
\label{sec:introduction}

Multilingual zero-shot TTS aims to synthesize natural speech for an unseen speaker from a short reference audio clip without further training~\cite{seedtts,f5tts,cosyvoice2,naturalspeech3,naturalspeech2,vall-e}. Recent progress in large-scale speech generation has made this setting increasingly practical, enabling speaker-consistent synthesis across languages for applications such as video dubbing, audiobooks, accessibility tools, and interactive voice assistants. In cross-lingual scenarios, the system must preserve the reference speaker's identity while generating intelligible and natural speech in a different target language.

Many zero-shot TTS systems condition on a paired text--audio prompt~\cite{seedtts,cosyvoice3,cosyvoice2,vall-e,omnivoice}, and therefore require a transcript of the reference audio at inference time. Accurate transcripts are unavailable for much in-the-wild speech, especially for low-resource languages and dialects~\cite{crosslingualf5tts,xvoice}. Several approaches have been explored to remove this dependency~\cite{rtfreef5,mosstts,crosslingualf5tts,xvoice,voxcpm2}, the most common of which uses a speaker encoder~\cite{tortoisetts,xtts,yourtts,indextts,qwen3tts,indextts25,chatterboxtts2025,minimaxtts,indextts2} to extract timbre features from the reference audio. Recent work supports both conditioning modes within a single model, and reports higher speaker similarity when the reference audio is used as a prefix~\cite{mosstts,voxcpm2}.

In this work, we present \sysname{}, a multilingual zero-shot TTS system. The model is trained across 14 languages and supports voice cloning in intra-lingual and cross-lingual settings. \sysname{} follows a two-stage architecture, consisting of text-to-semantic (T2S) and semantic-to-acoustic (S2A) modules. T2S predicts semantic speech tokens from the text and the speaker condition, and S2A renders them as mel-spectrograms with conditional flow matching~\cite{cfm} and a Diffusion Transformer (DiT) backbone~\cite{dit}. The T2S module uses a jointly trained speaker encoder, which extracts timbre features from self-supervised speech representations. At inference time, the same model also supports continuation cloning by conditioning on the reference transcript and the speech tokens of the reference audio.

Experimental results show that \sysname{} achieves top-tier cross-lingual intelligibility on CV3-Eval and remains among the leading systems on intra-lingual and multilingual zero-shot benchmarks. On our internal cross-lingual set, it obtains the best average overall rank in human evaluation among strong baseline systems. Our main contributions are as follows:
\begin{itemize}[leftmargin=*,itemsep=2pt,topsep=2pt,parsep=0pt]
    \item \sysname{}, a multilingual zero-shot TTS system that clones an unseen speaker in 14 languages without a reference transcript.

    \item A jointly trained speaker encoder that encodes reference audio into a speaker embedding for the T2S module.

    \item Two inference recipes with the same model, enabling transcript-free reference cloning by default and continuation cloning when a reference transcript is available.

    \item Objective and subjective evaluations against open-source and closed-source systems, on which \sysname{} performs competitively, and an open-source release of code, model checkpoints, and demos.
\end{itemize}

\section{Related Work}

\subsection{Cross-Lingual Zero-Shot TTS}

Zero-shot TTS clones the voice of an unseen speaker from a short audio prompt. Extending this capability across languages requires the synthesized utterance to remain intelligible and natural in a language different from that of the reference audio, while preserving the reference speaker's identity.

Many systems condition on the reference transcript together with the speech tokens of the reference audio as a paired text--audio prompt~\cite{seedtts,cosyvoice3,cosyvoice2,vall-e,omnivoice}. Recent work removes this dependency by leveraging forced alignment, a dedicated training strategy, or self-supervised representations of the reference audio~\cite{rtfreef5,crosslingualf5tts,xvoice}. Cross-Lingual F5-TTS~\cite{crosslingualf5tts} retains the flow-matching architecture and enables transcript-free audio prompting through Massively Multilingual Speech (MMS) forced alignment. At each training step, it randomly selects a word boundary, uses the preceding audio segment as the prompt, and discards its corresponding transcript. At inference time, it estimates the speaking rate from the audio prompt with a phoneme-, syllable-, or word-level predictor, and the target duration is the ratio of the number of linguistic units in the target text to the predicted speaking rate. X-Voice~\cite{xvoice} instead avoids forced alignment by training a multilingual flow-matching backbone, synthesizing speaker-consistent audio prompts, and fine-tuning on those synthetic pairs with the reference transcript masked, yielding transcript-free cloning without auxiliary modules. In contrast, \sysname{} removes the transcript from the conditioning signal: a jointly trained speaker encoder extracts the speaker condition from the reference audio, without forced alignment or synthetic prompt pairs.

\subsection{Reference Conditioning for Voice Cloning}

Reference conditioning can be broadly categorized into three types based on how the reference audio is represented.

The first type relies on paired text--audio prompts. VALL-E~\cite{vall-e} established this approach: the reference transcript is concatenated with the target text, the speech tokens of the reference serve as the prefix of the token sequence to be generated, and the model generates the target utterance as a continuation. Many in-context TTS systems build on this formulation~\cite{seedtts,cosyvoice,cosyvoice3,cosyvoice2,omnivoice}. This formulation retains fine-grained timbre, prosody, and style, but requires a reference transcript at inference time.

The second approach employs a global speaker embedding. ECAPA-TDNN~\cite{ecapatdnn} and CAM++~\cite{campplus} are representative speaker verification encoders, and YourTTS~\cite{yourtts} uses such an encoder for multilingual synthesis. Tortoise-TTS~\cite{tortoisetts} introduced the alternative of training the speaker encoder jointly with the generator, so that the resulting representation is tailored to the synthesis task~\cite{qwen3tts,minimaxtts}. XTTS~\cite{xtts}, Chatterbox-TTS~\cite{chatterboxtts2025}, the IndexTTS series~\cite{indextts,indextts25,indextts2}, Qwen3-TTS~\cite{qwen3tts}, and MiniMax-Speech~\cite{minimaxtts} follow this design. Unlike the first type, it does not require the transcript of the reference audio.

The third keeps the reference as a sequence of conditioning tokens without pairing it to a transcript. MOSS-TTS~\cite{mosstts} uses an optional speech prompt whose audio-token representations are concatenated with text embeddings. VoxCPM2~\cite{voxcpm2} encodes reference audio as a delimited prefix segment. This design retains fine-grained reference information and removes the transcript requirement~\cite{mosstts,voxcpm2}. \sysname{} uses the second design as the default reference-cloning mode and also supports the first design as a continuation-cloning inference mode when a reference transcript is available.

\section{Proposed Method}
\label{sec:method}

\subsection{Overview}

\begin{figure}[t]
  \centering
  \includegraphics[width=0.6\textwidth]{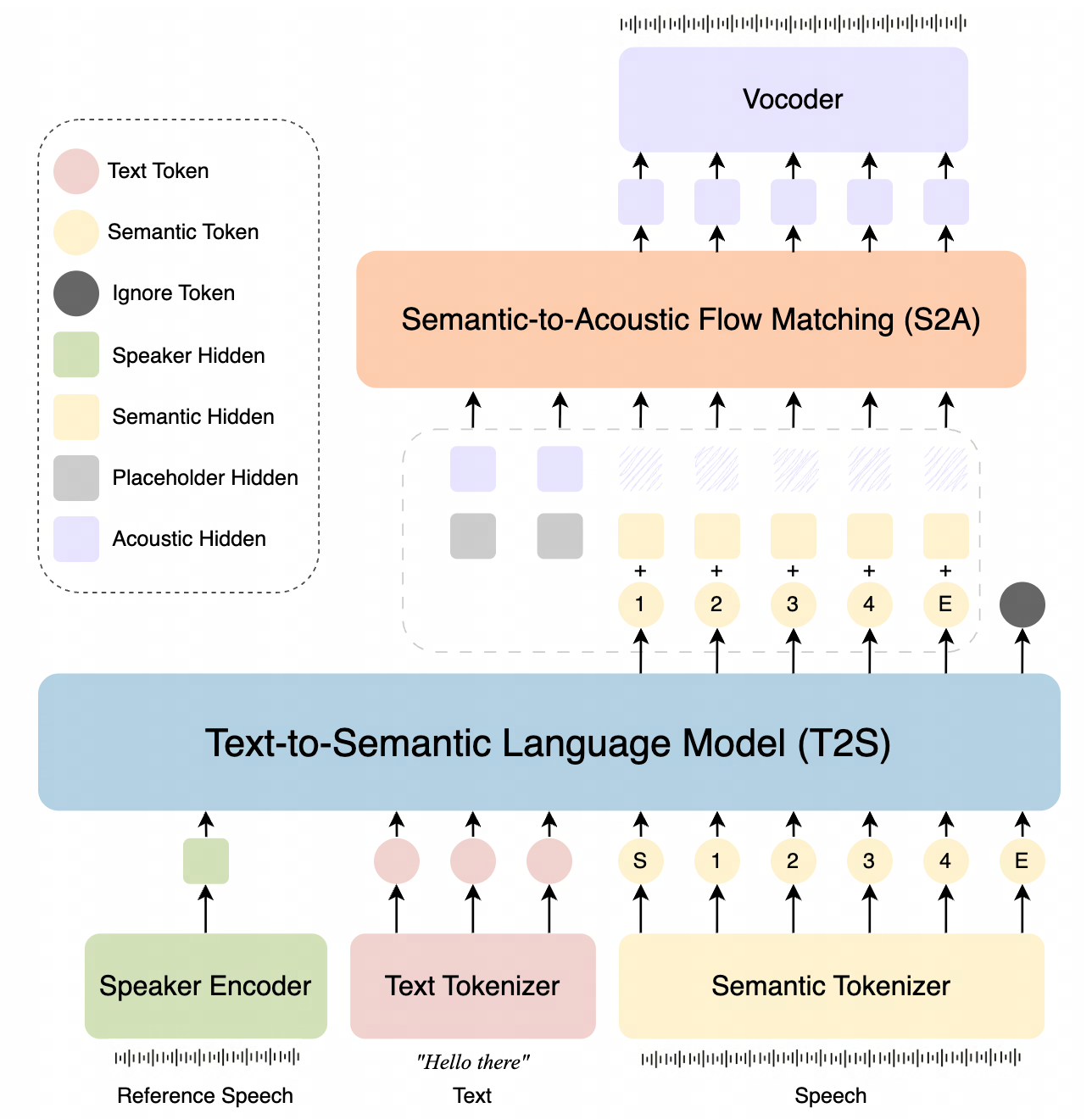}
  \caption{Overall architecture of \sysname{}, comprising the text-to-semantic module, the semantic-to-acoustic module, and the vocoder. The figure shows the training flow from reference audio and target text to semantic tokens and mel-spectrograms; the T2S inference layouts are given in Table~\ref{tab:layouts}.}
  \label{fig:overview}
\end{figure}

\sysname{} is a two-stage zero-shot TTS system designed for multilingual and cross-lingual voice cloning from untranscribed reference audio. As shown in Figure~\ref{fig:overview}, it comprises three components: an autoregressive text-to-semantic (T2S) module, a semantic-to-acoustic (S2A) module based on conditional flow matching (CFM)~\cite{voicebox,cfm,matchatts}, and a neural vocoder~\cite{bigvgan}. The T2S module predicts semantic tokens conditioned on the target text and the speaker embedding, which a learnable speaker encoder extracts from the reference audio. The S2A module predicts a mel-spectrogram from these predicted semantic tokens, the T2S hidden states, a global speaker embedding, and a prompt mel-spectrogram. The vocoder then converts the mel-spectrogram into a waveform.

\subsection{Text-to-Semantic Modeling}
\label{sec:t2s}

The T2S model is a decoder-only Transformer~\cite{gpt2}, and its training input sequence is constructed as
\begin{equation}
    [\,e^{r},\,H^{\mathrm{txt}},\,\langle\mathrm{BOS}\rangle,\,
    E^{\mathrm{T2S}}_{\mathrm{sem}}(y_1),\ldots,
    E^{\mathrm{T2S}}_{\mathrm{sem}}(y_N)\,],
\end{equation}
where $e^{r}$ is a speaker embedding pooled from the reference audio, $H^{\mathrm{txt}}$ denotes the target-text embeddings, $\langle\mathrm{BOS}\rangle$ is a boundary token separating text from speech, $E^{\mathrm{T2S}}_{\mathrm{sem}}$ denotes the T2S semantic-token embedding table, and $y=(y_1,\ldots,y_N)$ is the semantic-token sequence.

\paragraph{Text representation and language control.}
The target language is specified through the text input. Following the natural-language instruction format of the CosyVoice series~\cite{cosyvoice,cosyvoice3,cosyvoice2}, we prepend a short natural-language instruction before the input text for speech synthesis and denote the resulting target-text sequence by $x$. Text is processed using a BPE-based tokenizer taken from a pre-trained large language model~\cite{mistral7b}. No grapheme-to-phoneme module is required. The corresponding pre-trained embedding table $E^{\mathrm{LLM}}_{\mathrm{txt}}$ is kept frozen, and a lightweight learnable MLP $\psi_{\mathrm{txt}}$~\cite{qwen3tts} projects the pre-trained text embeddings to the T2S hidden dimension $d$, giving $H^{\mathrm{txt}} = \psi_{\mathrm{txt}}\big(E^{\mathrm{LLM}}_{\mathrm{txt}}(x)\big)$.

\paragraph{Speaker conditioning.}
The speaker encoder extracts timbre features from the reference audio $a^{r}$, transforming a variable-length reference into a fixed-size conditional vector. Following recent systems~\cite{qwen3tts,minimaxtts}, instead of using a speaker encoder pre-trained for the speaker verification task, we train the speaker encoder jointly with the T2S module so that the resulting representation matches the requirements of the synthesis task. In addition, the speaker encoder operates on self-supervised speech representations (SSL) rather than raw mel-spectrograms. SSL representations capture both content and speaker information~\cite{wavlm,w2vbert}, enabling the speaker encoder to extract the speaker embedding from the reference audio without its transcript~\cite{rtfreef5,indextts2}. The SSL encoder $E_{\mathrm{ssl}}$ first produces a frame-level representation $H^{\mathrm{ssl}} = E_{\mathrm{ssl}}(a^{r}) \in \mathbb{R}^{T_{\mathrm{ref}} \times D_{\mathrm{ssl}}}$, which the speaker encoder $f_{\mathrm{spk}}$ pools into a speaker embedding $e^{r}$:
\begin{equation}
    e^{r} = f_{\mathrm{spk}}\big(E_{\mathrm{ssl}}(a^{r})\big) \in \mathbb{R}^{d}.
\end{equation}
We adopt w2v-BERT~2.0~\cite{w2vbert} as $E_{\mathrm{ssl}}$. Since we set the output dimension of $f_{\mathrm{spk}}$ to the T2S hidden dimension $d$, the embedding $e^{r}$ can be prepended directly to the input sequence without an additional projection.

Following recent work~\cite{qwen3tts}, we adopt an ECAPA-TDNN architecture~\cite{ecapatdnn} as the speaker encoder, which aggregates temporal information through attentive statistics pooling~\cite{asp}.

\paragraph{Training and inference.}
We organize the training data by speaker. Each speaker has at least two utterances, and we construct each training pair by sampling two distinct utterances from the same speaker, one serving as the reference and the other as the target. Semantic tokens are extracted from the target speech $a$ by a frozen pre-trained semantic codec~\cite{maskgct}, with $y_i\in\{1,\ldots,V_{\mathrm{sem}}\}$. We append an end-of-sequence token $y_{N+1}=\langle\mathrm{EOS}\rangle$. The T2S model, parameterized by $\theta$, predicts each token from the target-text sequence $x$ and the speaker embedding $e^{r}$, and is trained with the autoregressive cross-entropy objective
\begin{equation}
    \mathcal{L}_{\mathrm{T2S}}
    =
    -\sum_{i=1}^{N+1}
    \log p_{\theta}(y_i \mid x, e^{r}, y_{<i}).
\end{equation}
We jointly optimize the speaker encoder, the text projection $\psi_{\mathrm{txt}}$, the semantic-token embedding, and the Transformer backbone, while the pre-trained text embedding table remains frozen. In addition to the generated tokens, we retain the T2S hidden states at the semantic-token positions, denoted by $H^{\mathrm{T2S}}$, and pass them to S2A. Conditioning the acoustic decoder on continuous hidden states rather than discrete tokens alone reduces the information bottleneck introduced by discrete quantization, and preserves context that is useful for content realization, prosody, and timbre~\cite{indextts,qwenaudio3tts,indextts2}.

\subsection{Dual-Mode Conditioning}
\label{sec:dual-mode}

The T2S model is trained in the reference-cloning layout. At inference time, the same model also supports continuation cloning by prepending the transcript and the speech tokens of the reference audio. Table~\ref{tab:layouts} shows the input sequence for each mode.

\begin{table}[!htbp]
\centering
\caption{T2S conditioning layouts of the two cloning modes. ``$\rightarrow$'' separates the conditioning input from the semantic tokens to be generated. $e^{r}$ is the pooled speaker embedding, $H^{\mathrm{txt},r}$ and $H^{\mathrm{txt}}$ are the reference-transcript and target-text embeddings, and $y^{r}$ denotes the reference semantic tokens.}
\label{tab:layouts}
\small
\begin{tabular}{ll}
\toprule
Mode & T2S conditioning layout \\
\midrule
Reference cloning    & $[\,e^{r}\,;\, H^{\mathrm{txt}}\,;\, \langle\mathrm{BOS}\rangle\,] \;\rightarrow\; y$ \\
Continuation cloning & $[\,e^{r}\,;\, H^{\mathrm{txt},r}\,;\, H^{\mathrm{txt}}\,;\, \langle\mathrm{BOS}\rangle\,;\, y^{r}\,] \;\rightarrow\; y$ \\
\bottomrule
\end{tabular}
\end{table}

In reference cloning, the T2S input contains only the speaker embedding and the target text. Generation is therefore not constrained by the reference prosody, allowing greater flexibility in prosody and style~\cite{minimaxtts}. Continuation cloning additionally conditions on the reference transcript and the reference semantic tokens, which yields higher speaker similarity~\cite{mosstts,voxcpm2}.

\subsection{Semantic-to-Acoustic Modeling}
\label{sec:s2a}

\paragraph{Acoustic conditioning.}
The S2A model converts the semantic sequence $y$ into a target mel-spectrogram $m$ with conditional flow matching and a Diffusion Transformer (DiT) backbone~\cite{dit}. During training, semantic tokens are extracted from the target speech by the same codec used for T2S; during inference, they are predicted by T2S. The semantic conditioning sequence combines semantic-token embeddings and T2S hidden states:
\begin{equation}
    c^{\mathrm{sem}} =
    \big[\,
    E^{\mathrm{S2A}}_{\mathrm{sem}}(y)\,;\,
    H^{\mathrm{T2S}}
    \,\big].
\end{equation}
The semantic conditioning sequence $c^{\mathrm{sem}}$ is upsampled to the mel-frame rate by a length regulator~\cite{seedvc}. A frozen speaker verification model $\mathrm{SV}$~\cite{ecapatdnn,campplus} produces a global speaker embedding $g$, which is repeated to the length of the upsampled sequence and concatenated with it along the feature dimension, giving the frame-level condition $c$. During training, $g=\mathrm{SV}(a)$ is extracted from the target utterance being reconstructed; during inference, $g^{r}=\mathrm{SV}(a^{r})$ is extracted from the reference audio.

S2A further receives a prompt mel-spectrogram $m^{p}$ as acoustic context~\cite{cosyvoice2,seedvc,minimaxtts,indextts2}. We prepend the prompt to the target along the time axis and replace the semantic conditioning over the prompt frames with a learnable placeholder embedding~\cite{xvoice}. The prompt therefore supplies acoustic evidence without providing explicit transcript- or token-level semantic conditioning for the prompt frames. During training, a randomly selected prefix of the target mel-spectrogram serves as the prompt and is excluded from loss calculation~\cite{seedvc}. During inference, $m^{p}=\mathrm{Mel}(a^{r})$ is derived from the reference audio.

\paragraph{Flow matching and classifier-free guidance.}
S2A is trained with conditional flow matching along an optimal transport path~\cite{cosyvoice2,cfm}, which matches the vector field $\omega_t$:
\begin{align}
    \phi^{\mathrm{OT}}_t(m_0,m_1) &= (1-t)\,m_0 + t\,m_1, \\
    \omega_t\big(\phi^{\mathrm{OT}}_t(m_0,m_1)\mid m_1\big) &= m_1 - m_0,
\end{align}
where $m_1\sim q(m)$ is a target mel-spectrogram drawn from the data distribution and $m_0\sim p_0(m)=\mathcal{N}(0,I)$ is a noise sample of the same shape. The DiT backbone, parameterized by $\phi$, estimates this vector field from the noisy input $\phi^{\mathrm{OT}}_t(m_0,m_1)$, the timestep $t$, the frame-level condition $c$, and the prompt mel-spectrogram $m^{p}$. It is optimized by minimizing the L1 loss between the predicted and the ground-truth vector field~\cite{cosyvoice2,seedvc,indextts2}:
\begin{equation}
    \mathcal{L}_{\mathrm{S2A}}
    =
    \mathbb{E}_{p_0(m),\,q(m),\,t}
    \left|
    \omega_t\big(\phi^{\mathrm{OT}}_t(m_0,m_1)\big)
    -
    v_{\phi}\big(\phi^{\mathrm{OT}}_t(m_0,m_1),\,t,\,c,\,m^{p}\big)
    \right|_1 .
\end{equation}
During training, the timestep follows a uniform distribution $\mathcal{U}[0,1]$.
To enable classifier-free guidance (CFG)~\cite{cfg} during inference, we also train the model on both conditional and unconditional situations---the semantic conditioning sequence, the speaker embedding, and the prompt mel are dropped jointly with a fixed probability:
\begin{equation}
\begin{aligned}
    \tilde{v}_{\phi}\big(\phi^{\mathrm{OT}}_t(m_0,m_1),\,t,\,c,\,m^{p}\big)
    &=
    (1+\alpha)\cdot
    v_{\phi}\big(\phi^{\mathrm{OT}}_t(m_0,m_1),\,t,\,c,\,m^{p}\big)
    \\
    &\quad -
    \alpha\cdot
    v_{\phi}\big(\phi^{\mathrm{OT}}_t(m_0,m_1),\,t,\,\emptyset,\,\emptyset\big),
\end{aligned}
\end{equation}
where $\alpha$ is the CFG strength and $\alpha=0$ recovers the purely conditional field. During inference, we integrate the guided vector field from noise to data with the Euler ODE solver over a uniform time discretization of $[0,1]$, and reconstruct the waveform with a pre-trained neural vocoder~\cite{bigvgan}.

\section{Experiments}
\label{sec:experiments}

\subsection{Experimental Setup}

\paragraph{Training data.}
\sysname{} is trained on approximately 500k hours of multilingual speech covering 14 languages: Chinese, English, Japanese, Korean, German, French, Spanish, Indonesian, Italian, Thai, Portuguese, Russian, Malay, and Vietnamese. The corpus comprises real and synthetic speech. We curate the training corpus with a data processing pipeline. The pipeline consists of source separation and denoising, voice-activity-detection-based segmentation, filtering out multi-speaker, overlapping, and low-quality segments, language identification, multi-system ASR filtering, and speaker clustering. We retain utterances whose cross-model ASR error rate is below $2.5\%$ (character error rate for Chinese, Japanese, and Korean, and word error rate for all other languages). We additionally include roughly 1,000 hours of synthetic speech per language to cover very short utterances and patterns underrepresented in real speech; this synthetic portion accounts for a small portion of the overall training corpus. Language-based resampling is used to balance the sampling weights across high- and low-resource languages. For speaker conditioning, the reference and target are different recordings of the same speaker.

\paragraph{Model configuration.}
The T2S model is a 24-layer decoder-only Transformer with a hidden size of 1280. It uses a pre-trained LLM tokenizer~\cite{mistral7b}; the embedding table is projected to the T2S hidden dimension by a lightweight MLP. An ECAPA-TDNN speaker encoder~\cite{ecapatdnn,qwen3tts} produces a speaker embedding from the features of a w2v-BERT~2.0~\cite{w2vbert} encoder. Semantic tokens are extracted with a pre-trained MaskGCT~\cite{maskgct} semantic tokenizer. The S2A model is a conditional flow-matching decoder with a Diffusion Transformer~\cite{dit} backbone. It renders 80-dimensional mel-spectrograms from semantic tokens, T2S hidden states, a global embedding from a CAM++~\cite{campplus} speaker verification model, and a reference mel-spectrogram prompt; a pre-trained BigVGAN~\cite{bigvgan} vocoder reconstructs the waveform.

\paragraph{Training and inference.}
\sysname{} is trained in two stages on 32 NVIDIA A40 GPUs. First, the T2S model and learnable speaker encoder are jointly optimized for autoregressive semantic-token prediction. The pre-trained LLM embedding table remains frozen. Second, we freeze T2S and train S2A with conditional flow matching. Condition dropout during S2A training enables classifier-free guidance at inference. Both stages use AdamW~\cite{adamw} with a cosine learning-rate schedule.

At inference time, \sysname{} requires only target text and a short reference audio. The reference is converted into a speaker embedding for T2S, a global speaker embedding for S2A, and a reference mel-spectrogram prompt. T2S generates semantic tokens autoregressively; S2A then generates a mel-spectrogram with 25 Euler steps and classifier-free guidance (guidance strength $\alpha = 0.7$). Finally, the vocoder converts the mel-spectrogram to a waveform. Continuation-cloning results are explicitly marked in the tables.

\paragraph{Benchmarks.}
We evaluate \sysname{} on four public benchmarks. CV3-Eval~\cite{cosyvoice3,cv3eval} is an in-the-wild multilingual voice cloning benchmark released with CosyVoice~3, built on reference speech from Common Voice, FLEURS, and web-crawled recordings; we evaluate on its cross-lingual subset. X-Voice~\cite{xvoice} covers 30 languages with human-recorded utterances drawn mainly from Common Voice, and we evaluate the cross-lingual pairs whose target text is Chinese. Seed-TTS-eval~\cite{seedtts} is a widely used Chinese--English zero-shot voice cloning benchmark. MiniMax-MLS-Test~\cite{minimaxtts} covers 24 languages, with 100 utterances per language and two Common Voice reference speakers per language.

\paragraph{Metrics.}
We report word error rate (WER) or character error rate (CER) for intelligibility and speaker similarity (SIM) for speaker preservation. CER is used for Chinese, Japanese, and Korean, and WER for all other languages. Both are computed using Whisper large-v3~\cite{whisper} for non-Chinese languages and a Paraformer~\cite{paraformer} ASR model for Chinese. SIM is the cosine similarity between speaker embeddings of the generated and reference audio, extracted with the fine-tuned WavLM-large~\cite{wavlm} speaker verification model used in Seed-TTS-eval~\cite{seedtts}.

\paragraph{Baselines.}
We compare \sysname{} with recent open-source releases~\cite{cosyvoice3,cosyvoice2,fishaudio,qwen3tts,xvoice,voxcpm2,omnivoice} and commercial systems~\cite{elevenlabs,minimaxtts}. Evaluated configurations that use a reference transcript at inference time are marked with $\dagger$. For systems supporting multiple cloning modes, this marker refers to the configuration evaluated here rather than to the full capability of the system.

\subsection{Cross-Lingual Voice Cloning}
\label{sec:cross-lingual}

\paragraph{CV3-Eval.}
We evaluate the six directions among Chinese, English, Japanese, and Korean that target Chinese or English, with 200 utterances per direction (1,200 in total). Table~\ref{tab:cv3} reports cross-lingual WER/CER on CV3-Eval across six source--target language pairs. \sysname{} achieves the lowest error rate on four of the six directions and remains close to the best system on the other two. Its margin over CosyVoice~2 is especially large for Japanese and Korean references (e.g., 4.87 vs.\ 48.10 for ja$\to$zh).

\begin{table}[!htbp]
\centering
\caption{Cross-lingual WER/CER (\%, $\downarrow$) on CV3-Eval. Chinese targets are scored with CER and English targets with WER. $\dagger$ denotes evaluated configurations that use a reference transcript at inference time. Boldface marks the best result per row.}
\label{tab:cv3}
\small
\setlength{\tabcolsep}{4pt}
\begin{tabular}{lrrrrrr}
\toprule
Direction & \textbf{Ours} & CosyVoice~2$^\dagger$ & CosyVoice~3-0.5B$^\dagger$ & CosyVoice~3-1.5B$^\dagger$ & OmniVoice$^\dagger$ & VoxCPM2 \\
\midrule
en$\to$zh & \textbf{6.16} & 13.50 & 8.48 & 8.01 & 6.53 & 6.29 \\
ja$\to$zh & 4.87 & 48.10 & 6.86 & 6.78 & 52.64 & \textbf{4.20} \\
ko$\to$zh & 1.28 & 7.70  & 5.24 & 3.30 & 1.71 & \textbf{1.20} \\
zh$\to$en & \textbf{3.19} & 17.10 & 6.83 & 5.39 & 3.72 & 3.84 \\
ja$\to$en & \textbf{3.44} & 11.20 & 5.86 & 5.94 & 5.25 & 4.10 \\
ko$\to$en & \textbf{3.42} & 13.10 & 18.30 & 13.70 & 3.91 & 5.69 \\
\bottomrule
\end{tabular}
\end{table}

\paragraph{X-Voice.}
We evaluate the seven X-Voice source languages that \sysname{} supports---German, English, French, Japanese, Korean, Thai, and Vietnamese---with 500 utterances per direction (3,500 in total). Table~\ref{tab:xvoice} reports CER across seven source-to-Chinese directions. \sysname{} obtains the lowest CER on four directions (de$\to$zh, fr$\to$zh, ko$\to$zh, and vi$\to$zh) and is within 0.3 absolute points of the best system on the remaining directions.

\begin{table}[!htbp]
\centering
\caption{Cross-lingual CER (\%, $\downarrow$) on X-Voice (source$\to$zh). $\dagger$ denotes evaluated configurations that use a reference transcript at inference time. Boldface marks the best result per row.}
\label{tab:xvoice}
\small
\begin{tabular}{lrrrrr}
\toprule
Direction & \textbf{Ours} & X-Voice & IndexTTS2 & OmniVoice$^\dagger$ & VoxCPM2 \\
\midrule
de$\to$zh & \textbf{2.86} & 3.07 & 3.46 & 7.79 & 3.62 \\
en$\to$zh & 3.21 & \textbf{3.06} & 3.78 & 3.30 & 3.35 \\
fr$\to$zh & \textbf{2.70} & 3.01 & 3.53 & 8.16 & 3.75 \\
ja$\to$zh & 3.50 & \textbf{3.39} & 4.11 & 60.88 & 4.53 \\
ko$\to$zh & \textbf{2.86} & 3.13 & 2.90 & 7.35 & 6.33 \\
th$\to$zh & 2.82 & \textbf{2.79} & 3.08 & 2.85 & 5.96 \\
vi$\to$zh & \textbf{2.75} & 2.78 & 2.98 & 6.59 & 3.65 \\
\bottomrule
\end{tabular}
\end{table}

\subsection{Intra-Lingual Voice Cloning}
\label{sec:intra-lingual}

The \texttt{test-en} subset of Seed-TTS-eval contains 1,088 samples drawn from Common Voice, and the \texttt{test-zh} subset contains 2,020 samples drawn from DiDiSpeech; each sample pairs a reference audio with a target sentence in the same language. Table~\ref{tab:seed} reports intra-lingual zero-shot results for both reference cloning and continuation cloning. Reference cloning achieves 1.49 WER and 0.700 SIM on English, and 0.94 CER and 0.765 SIM on Chinese. Continuation cloning improves speaker similarity on English (0.700 to 0.715) when a reference transcript is available, with a higher WER than reference cloning (1.68 vs.\ 1.49). Compared with strong baselines, reference cloning remains competitive in intelligibility, while Seed-TTS obtains the highest speaker similarity on both English and Chinese.

\begin{table}[!htbp]
\centering
\caption{Intra-lingual zero-shot results on Seed-TTS-eval. $\dagger$ denotes evaluated configurations that use a reference transcript at inference time. Boldface marks the best result per column.}
\label{tab:seed}
\small
\setlength{\tabcolsep}{5pt}
\begin{tabular}{lcccc}
\toprule
\multirow{2}{*}{System} & \multicolumn{2}{c}{English} & \multicolumn{2}{c}{Chinese} \\
\cmidrule(lr){2-3} \cmidrule(lr){4-5}
 & WER (\%, $\downarrow$) & SIM ($\uparrow$) & CER (\%, $\downarrow$) & SIM ($\uparrow$) \\
\midrule
\textbf{Ours} & 1.49 & 0.700 & 0.94 & 0.765 \\
\textbf{Ours (Continuation)}$^\dagger$ & 1.68 & 0.715 & 1.15 & 0.766 \\
Seed-TTS$^\dagger$ & 2.25 & \textbf{0.762} & 1.12 & \textbf{0.796} \\
Qwen3-TTS$^\dagger$ & \textbf{1.24} & 0.714 & \textbf{0.77} & 0.770 \\
FishAudio S2$^\dagger$ & 1.79 & 0.643 & 0.98 & 0.737 \\
OmniVoice$^\dagger$ & 1.62 & 0.740 & 0.87 & 0.777 \\
VoxCPM2$^\dagger$ & 1.70 & 0.752 & 0.97 & 0.793 \\
X-Voice & 1.91 & 0.627 & 1.47 & 0.746 \\
\bottomrule
\end{tabular}
\end{table}

\subsection{Multilingual Voice Cloning}
\label{sec:multilingual}

MiniMax-MLS-Test provides 100 target sentences and two Common Voice reference speakers per language, one female and one male, with 50 sentences assigned to each speaker. Following the practice of reporting only the languages a system supports~\cite{xvoice}, we evaluate the 11 languages that are both present in this test set and among the 14 languages \sysname{} covers. Table~\ref{tab:minimax} compares reference cloning with continuation cloning, as well as prior systems, across 11 languages. Reference cloning obtains the lowest WER on German and Thai and remains close to the best systems on Indonesian, Korean, Italian, and Spanish. Across all evaluated languages, \sysname{} achieves speaker similarity above 72\%. Continuation cloning consistently improves SIM across all 11 languages when a reference transcript is available.

\begin{table}[!htbp]
\centering
\caption{Per-language results on MiniMax-MLS-Test. Panel (a) reports WER/CER (\%, lower is better): Korean and Japanese are scored with CER, and the other languages with WER. Panel (b) reports SIM (higher is better). MiniMax-Speech and ElevenLabs numbers are taken from~\cite{minimaxtts}, where ElevenLabs denotes Eleven Multilingual v2. $\dagger$ denotes evaluated configurations that use a reference transcript at inference time. Boldface marks the best result per row.}
\label{tab:minimax}
\small
\setlength{\tabcolsep}{3.5pt}
\resizebox{\textwidth}{!}{%
\begin{tabular}{lrrrrrrrr}
\toprule
Language & \textbf{Ours} & \textbf{Ours (Cont.)}$^\dagger$ & MiniMax-Speech & ElevenLabs & Qwen3-TTS$^\dagger$ & FishAudio S2$^\dagger$ & OmniVoice$^\dagger$ & VoxCPM2$^\dagger$ \\
\midrule
\multicolumn{9}{c}{(a) Intelligibility: WER/CER (\%, $\downarrow$)} \\
\midrule
German     & \textbf{0.47} & 0.68 & 1.91 & 0.57 & 1.24 & 0.55 & 0.80 & 1.12 \\
French     & 3.66 & 4.87 & 4.10 & 5.22 & \textbf{2.86} & 3.90 & 3.58 & 3.42 \\
Indonesian & 1.12 & 1.41 & 1.24 & \textbf{1.06} & -- & 2.93 & 1.34 & 1.17 \\
Korean     & 1.84 & 2.50 & 1.75 & 1.87 & 1.76 & \textbf{1.62} & 2.66 & 3.34 \\
Thai       & \textbf{1.56} & 2.47 & 2.70 & 73.94 & -- & 6.66 & 2.93 & 2.19 \\
Japanese   & 4.14 & 4.05 & 3.52 & 10.65 & 3.82 & 3.52 & 3.59 & \textbf{3.51} \\
Vietnamese & 1.61 & 1.59 & \textbf{0.88} & 73.42 & -- & 14.11 & 0.95 & 4.19 \\
Italian    & 1.30 & 3.26 & 1.54 & 1.74 & \textbf{0.95} & 1.49 & 1.20 & 1.34 \\
Portuguese & 2.48 & 3.91 & 1.88 & \textbf{1.33} & 1.53 & 1.57 & 1.83 & 1.71 \\
Spanish    & 1.02 & 1.65 & 1.03 & 1.08 & 1.13 & 0.95 & \textbf{0.81} & 1.32 \\
Russian    & 4.64 & 5.42 & 4.28 & 3.88 & \textbf{3.21} & 4.24 & 4.63 & 4.53 \\
\midrule
\multicolumn{9}{c}{(b) Speaker similarity: SIM ($\uparrow$)} \\
\midrule
German     & 0.775 & 0.777 & 0.733 & 0.614 & 0.768 & 0.706 & 0.804 & \textbf{0.805} \\
French     & 0.723 & 0.755 & 0.628 & 0.535 & 0.716 & 0.658 & \textbf{0.776} & 0.738 \\
Indonesian & 0.765 & 0.767 & 0.729 & 0.660 & -- & 0.736 & 0.777 & \textbf{0.795} \\
Korean     & 0.812 & 0.824 & 0.776 & 0.700 & 0.790 & 0.742 & 0.831 & \textbf{0.837} \\
Thai       & 0.773 & 0.807 & 0.800 & 0.588 & -- & 0.749 & \textbf{0.847} & 0.841 \\
Japanese   & 0.788 & 0.806 & 0.776 & 0.738 & 0.771 & 0.753 & 0.821 & \textbf{0.825} \\
Vietnamese & 0.751 & 0.753 & 0.743 & 0.369 & -- & 0.693 & 0.775 & \textbf{0.793} \\
Italian    & 0.787 & 0.791 & 0.699 & 0.579 & 0.752 & 0.764 & \textbf{0.813} & 0.779 \\
Portuguese & 0.796 & 0.801 & 0.805 & 0.711 & 0.805 & 0.777 & \textbf{0.866} & 0.842 \\
Spanish    & 0.778 & 0.794 & 0.762 & 0.615 & 0.814 & 0.734 & 0.814 & \textbf{0.829} \\
Russian    & 0.787 & 0.796 & 0.761 & 0.675 & 0.784 & 0.768 & 0.784 & \textbf{0.807} \\
\bottomrule
\end{tabular}
}
\end{table}

\subsection{Human Evaluation}
\label{sec:human-eval}

To complement the automatic metrics above, we conduct a human evaluation of \sysname{} against the five baseline systems shown in Table~\ref{tab:human}, on an internal cross-lingual test set covering zh$\to$en, en$\to$zh, zh$\to$ko, and zh$\to$ja. Raters are native or fluent speakers of the target language of the direction they assess. For every test utterance, raters independently rank the six systems' outputs from 1 to 6, with 1 indicating the best system and ties averaged. The four dimensions are timbre similarity, naturalness, pronunciation authenticity, and emotional expressiveness; an overall rank is collected separately. Ratings are collected blind: system identities are hidden from the raters. We report the average rank per system, dimension, and language pair. Lower is better.

Table~\ref{tab:human} shows that \sysname{} ranks first or second in timbre similarity and naturalness on all four language pairs. It also achieves the best overall rank on en$\to$zh, zh$\to$ko, and zh$\to$ja; VoxCPM2 ranks first on zh$\to$en. On the remaining two dimensions, \sysname{} ranks first or second in pronunciation authenticity on all four language pairs, and first or second in emotional expressiveness on three of the four.

\begin{table}[!htbp]
\centering
\caption{Human evaluation: average rank (1--6, lower is better) averaged over raters. MiniMax-Speech denotes MiniMax-Speech-2.8-HD, and ElevenLabs denotes Eleven v3. $\dagger$ denotes evaluated configurations that use a reference transcript at inference time. Boldface marks the best result per row.}
\label{tab:human}
\small
\setlength{\tabcolsep}{4pt}
\resizebox{\textwidth}{!}{%
\begin{tabular}{llrrrrrr}
\toprule
Direction & Dimension & \textbf{Ours} & ElevenLabs & MiniMax-Speech & OmniVoice$^\dagger$ & Qwen3-TTS$^\dagger$ & VoxCPM2$^\dagger$ \\
\midrule
\multirow{5}{*}{zh$\to$en} & Timbre similarity & \textbf{1.60} & 3.38 & 3.63 & 2.50 & 3.70 & 2.68 \\
 & Naturalness & \textbf{2.38} & 2.60 & 3.28 & 3.03 & 2.95 & 2.58 \\
 & Pronunciation & 1.85 & 2.48 & 4.48 & 3.05 & 4.85 & \textbf{1.65} \\
 & Emotion & 2.95 & 1.80 & 3.08 & 3.13 & 3.25 & \textbf{1.75} \\
 & Overall & 2.28 & 2.95 & 4.15 & 3.75 & 4.43 & \textbf{1.95} \\
\midrule
\multirow{5}{*}{en$\to$zh} & Timbre similarity & \textbf{1.33} & 3.40 & 3.10 & 2.50 & 3.05 & 2.17 \\
 & Naturalness & 2.25 & \textbf{2.15} & 3.40 & 3.45 & 3.40 & 2.40 \\
 & Pronunciation & \textbf{1.80} & 2.10 & 4.12 & 2.73 & 4.00 & 1.88 \\
 & Emotion & 1.82 & \textbf{1.73} & 2.70 & 3.35 & 3.00 & 1.92 \\
 & Overall & \textbf{1.97} & 2.62 & 4.17 & 3.63 & 4.28 & 2.16 \\
\midrule
\multirow{5}{*}{zh$\to$ko} & Timbre similarity & \textbf{1.55} & 5.60 & 3.00 & 2.25 & 4.85 & 3.75 \\
 & Naturalness & \textbf{1.75} & 2.70 & 4.25 & 3.10 & 5.35 & 3.85 \\
 & Pronunciation & 2.15 & \textbf{1.30} & 3.75 & 4.50 & 5.85 & 3.25 \\
 & Emotion & \textbf{2.15} & 4.75 & 2.50 & 3.05 & 5.50 & 3.05 \\
 & Overall & \textbf{1.20} & 3.65 & 2.95 & 3.55 & 5.85 & 3.80 \\
\midrule
\multirow{5}{*}{zh$\to$ja} & Timbre similarity & \textbf{1.55} & 4.25 & 3.35 & 3.00 & 4.75 & 4.10 \\
 & Naturalness & \textbf{2.05} & 2.70 & 4.15 & 2.25 & 5.65 & 4.20 \\
 & Pronunciation & 2.05 & \textbf{1.65} & 3.90 & 3.20 & 5.80 & 4.25 \\
 & Emotion & \textbf{1.55} & 3.65 & 4.15 & 2.90 & 4.55 & 4.20 \\
 & Overall & \textbf{1.50} & 2.30 & 4.25 & 2.75 & 5.85 & 4.30 \\
\bottomrule
\end{tabular}
}
\end{table}

\section{Conclusion}

We presented \sysname{}, a multilingual zero-shot TTS system capable of cloning an unseen speaker across 14 languages from a short reference audio and target text, without requiring a reference transcript. When a reference transcript is available, continuation cloning further improves speaker similarity, at the cost of a slightly higher error rate. The system combines autoregressive text-to-semantic modeling, speaker conditioning from self-supervised speech representations, and a conditional flow-matching semantic-to-acoustic decoder. Objective and subjective evaluations show high intelligibility and speaker similarity across cross-lingual, intra-lingual, and multilingual settings. We release the code, model checkpoints, and demos to facilitate further research on multilingual and cross-lingual speech generation. Future work will focus on expanding high-quality long-tail language data and Chinese dialect coverage, improving streaming and low-latency inference, reducing runtime cost for production serving, and strengthening speaker similarity and fine-grained style preservation under diverse reference conditions.

\bibliographystyle{plain}
\bibliography{ref}





\end{document}